# COSMOLOGICAL SIMULATIONS USING SPECIAL PURPOSE COMPUTERS: IMPLEMENTING P³M ON GRAPE

Philippe P. Brieu, FJ Summers, and Jeremiah P. Ostriker
Princeton University Observatory, Peyton Hall, Ivy Lane, Princeton, NJ 08544–1001
I: philippe@Princeton.EDU, summers@grape.Princeton.EDU, jpo@astro.Princeton.EDU

## ABSTRACT

An adaptation of the Particle–Particle/Particle–Mesh (P³M) code to the special purpose hardware GRAPE is presented. The short range force is calculated by a four chip GRAPE–3A board, while the rest of the calculation is performed on a Sun SPARC 10/51 workstation. The limited precision of the GRAPE hardware and algorithm constraints introduce stochastic errors of the order of a few percent in the gravitational forces. Tests of this new P3MG3A code show that it is a robust tool for cosmological simulations. The code currently achieves a peak efficiency of one third the speed of the vectorized P³M code on a Cray C–90 and significant improvements are planned in the near future. Special purpose computers like GRAPE are therefore an attractive alternative to supercomputers for numerical cosmology.

*Subject headings:* cosmology: large–scale structure of universe — galaxies: clustering — methods: numerical

## 1. INTRODUCTION

During the past decade cosmologists have increasingly used numerical simulations to test various models of the evolution of large scale structures and the process of galaxy formation. Advances in cosmological simulations have followed the increase in computing power. Simulations are running faster and at a higher resolution as computer speed and memory have increased dramatically in the past few years, thereby allowing increases in the number of particles and mesh points used or the addition of hydrodynamics.

Most of these simulations have been performed on supercomputers. However, in the late 1980s, Junichiro Makino and his team at the University of Tokyo started developing a series of computer chip boards that could perform N–body type calculations at a speed comparable to that of supercomputers (Sugimoto et al. 1990) for a fraction of their cost (typically, one thousandth). These GRAPE boards (a contraction of "GRAvity piPE") are specially designed to compute forces on particles by an extremely fast direct summation method. The GRAPE project has generated several kinds of boards (see Ebisuzaki et al. 1993 for a review). Boards in the GRAPE–2 series have a higher accuracy, but are slower than the GRAPE–1 and GRAPE–3 series (see Table 1 of Ebisuzaki et al. 1993). GRAPE boards have been used for simulations of galaxy mergers (Ebisuzaki, Makino, & Okumura 1991), violent relaxation (Funato, Makino, & Ebisuzaki 1992), and other astrophysical problems such as the evolution of binary black holes (Makino et al. 1993) and clusters of galaxies (Funato, Makino, & Ebisuzaki 1993). Furthermore, they have been used for treecodes (Makino 1991) and Smoothed Particle Hydrodynamics (Umemura et al. 1993).

We are interested in large cosmological simulations, and GRAPE–3A is the best suited of all commercially available GRAPE boards for this kind of problem. The main gain expected from GRAPE is in the speed of computation of particle accelerations. Codes that will benefit most from a GRAPE implementation are those that use direct summation techniques the most heavily.

The P³M code (which stands for "Particle–Particle/Particle–Mesh," see Hockney & Eastwood 1981, hereafter HE) is a good choice because the Particle–Mesh (PM) calculation provides a fast long range force calculation with periodic boundary conditions while most of the computational effort is due to the local Particle–Particle (PP) calculations (direct summation). Combining GRAPE and P³M will enable us to do much larger simulations than previous pure direct summation codes using GRAPE. In addition, those simulations can be run on a local workstation at speeds approaching supercomputer versions of P³M.

In this paper, we give a brief description of GRAPE and its performance. We then explain how we implemented a P³M code on a GRAPE–3A board. Finally, we show tests of the code and results from a cosmological simulation.

## 2. GRAPE

### 2.1. *Technical Description*

We use a four chip GRAPE–3A board, which we will simply refer to as "GRAPE" or "the board" hereafter. The idea behind GRAPE is to shift most of the computational effort from a regular computer to the board, which performs a very limited, but highly efficient set of hard wired calculations. The board is connected to the host workstation by a bus as shown on Figure 1. In our case the host is a Sun SPARC 10/51 with an S bus connected to a Solflower VME cage which holds the GRAPE board. Only the short range force, potential, and neighbor list calculations are done by the board. All other parts of the simulation are handled by the host.

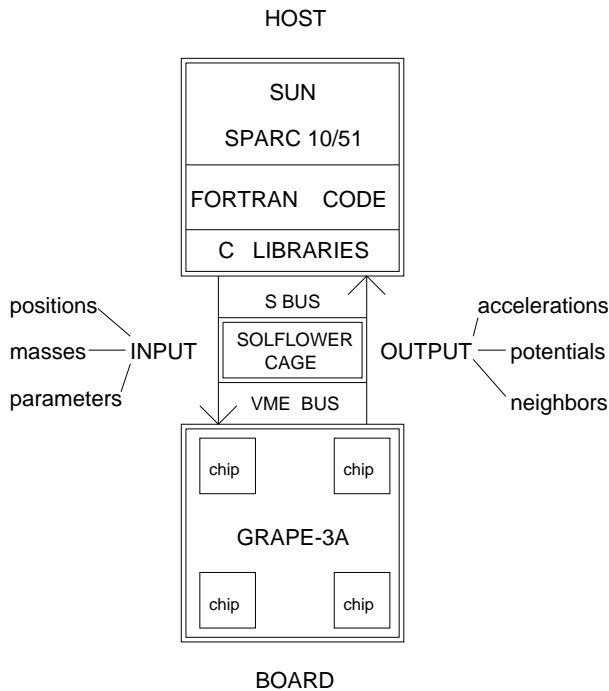

FIG. 1.—Schematic of the GRAPE hardware used, with simplified information flow charts.

A C language library (with FORTRAN wrappers) is the interface between the user's program and the board. The user loads in the positions and masses of the neighbor particles and sets their gravitational softening parameter (for the hard wired Plummer force law). Then one loops over home particles (often the same set as the neighbor particles), sends a home particle position and neighbor search radius, and gets back the accelerations, potential, and list of neighbors within the search radius for that particle. In practice, four home particles (one per available chip) are sent to the board, and the neighbor list returned is a concatenation of all four particles' lists.

The GRAPE board does not use the standard 32 bit floating point format to represent real numbers. for various reasons (Okumura et al. 1993, Fukushige et al. 1991) a variety of numerical representations are used in the hardware, including 20 bit and 56 bit fixed point formats and 13 bit unsigned logarithmic format. Conversions to and from these formats are handled by the board using mass and length scales set by the user at the beginning of the simulation. Particular care must be taken in setting these parameters to achieve maximum resolution without danger of unwanted underflows or overflows. The limitation on spatial resolution is set by the scaling of the input positions to integers in the range $[-2^{18}, 2^{18}] = \pm 262,144$.

This design has additional limitations. The maximum number of particles GRAPE can handle at a time is 32,768 and the maximum number of neighbors it can return is 256 (an eight chip version of the board raises these numbers to 131,072 and 1,024, respectively). Accelerations, potentials, and neighbor lists will have intrinsic errors due to roundoff in the format conversions (of the order of a few percent as discussed below). Also, underflows and overflows that occur during a calculation are silently set to zero (underflow) or the maximum value of the current range (overflow). An acceleration that would be larger than the maximum allowed (i.e. the value calculated with the maximum mass and minimum distance scales) is arbitrarily inaccurate.

### 2.2. Timing

Timings of the speed of a direct summation calculation using GRAPE–3A ($t_{G3A}$) for various numbers of particles ($N$) are given in Table 1. For very small numbers of particles ($N \leq 100$), the time required to compute the accelerations on the particles ($\simeq$ 2 ms) is independent of $N$ because the communication time between host and board dominates. As more and more particles are added to the calculation, the hardware speed becomes dominant. For very large $N$, the computation time scales as $5.3 \times 10^{-8} \times N^2/n_{chips}$ seconds (where $n_{chips}$ is the number of chips on the board — four in our case). From this scaling one can see that direct summation calculations using GRAPE are limited to $\sim 2.6 \times 10^5 \times \sqrt{n_{chips}}$ particles if we want to limit the CPU time to less than an hour per timestep.

Table 1 also provides a comparison to an optimized direct summation calculation ($t_{ODS}$) done on the host. The optimizations include taking advantage of force symmetry between particles and utilizing a lookup table for the force law. For more than $N \simeq 10,000$ particles, the GRAPE calculation is roughly 100 times faster than the optimized direct summation on the host. This is the maximum gain in speed to be expected from GRAPE for any code using direct summation and therefore the maximum one may expect for the PP part of the $P^3M$ code. In tests against a treecode with the maximum number of particles, depending on the parameters used (monopole or quadrupole, opening angle, etc.), we found GRAPE to be five to ten times faster than a treecode, which scales as $N \log N$ (Barnes & Hut 1986).

### 2.3. Numerical Performance

#### 2.3.1. Forces

Errors in forces are discussed in Makino, Ito, & Ebisuzaki (1990) for GRAPE–1 and Okumura et al. (1993) for GRAPE–3. These inaccuracies are due to the roundoff errors in positions and masses during scaling, internal conversions, and summation. For pairwise interactions averaged over random separations, the 3D error distribution is Gaussian with a typical r.m.s. of 2% (see

TABLE 1
CPU Time to Compute Particle Accelerations

| $N_{part}$ | $t_{G3A}$ (min.) | $t_{ODS}$ (min.) | $t_{ODS}/t_{G3A}$ |
|---|---|---|---|
| 100 | $2.2 \times 10^{-4}$ | $2.5 \times 10^{-4}$ | 1 |
| 300 | $4.0 \times 10^{-4}$ | $1.9 \times 10^{-3}$ | 5 |
| 1,000 | $6.8 \times 10^{-4}$ | $2.1 \times 10^{-2}$ | 31 |
| 3,000 | $3.2 \times 10^{-3}$ | $2.1 \times 10^{-1}$ | 66 |
| 10,000 | $2.5 \times 10^{-2}$ | 2.3 | 93 |
| 30,000 | $2.0 \times 10^{-1}$ | $2.0 \times 10^{1}$ | 100 |
| 100,000 | 2.2 | $2.2 \times 10^{2}$ | 99 |

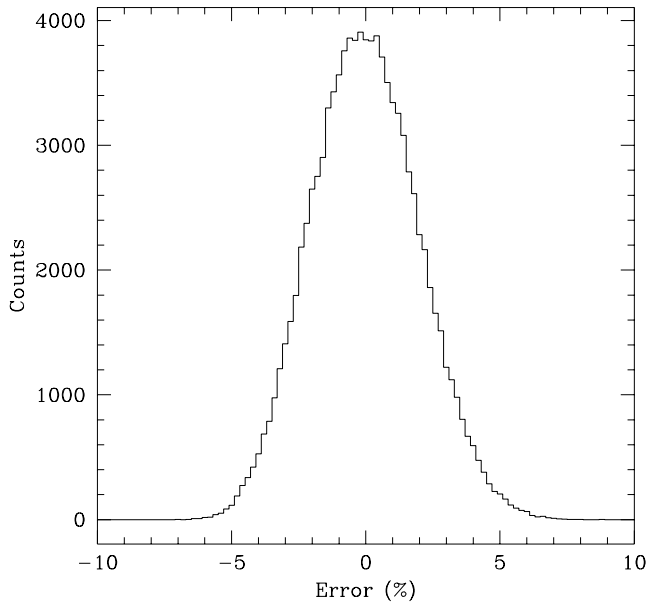

FIG. 2.—Histogram of the distribution of pairwise force errors in the total force returned by GRAPE for randomly separated pairs of particles.

Figure 2). At very small separations (i.e. much smaller than the softening parameter), roundoff and underflow can produce errors of up to 100%. At such scales, the force is already heavily softened and its contribution to the total force and dynamics is insignificant.

Table 2 shows how the r.m.s. force errors decrease with increasing number of particles ($N_{part}$). In each case, 100,000 calculations were made ten times to obtain a statistical set. Although the mean is systematically positive in this example, it depends on the choice of length scale and softening parameter in a somewhat chaotic way. In all cases of interest, the mean will be between $-0.2\%$ and $0.2\%$. Note also that Newton's second law does not apply exactly to the forces computed by GRAPE — regardless of the option to have particles with different softenings — as the interaction is computed twice for each pair of particles (once with each particle as home particle) and roundoff errors accrue independently. Therefore, linear momentum is not strictly conserved.

One aspect not discussed in the papers referenced above is an anisotropy in the errors. In calculating $\Delta r^2 + \epsilon^2$ for the acceleration and potential, the hardware utilizes the grouping $(\Delta x^2 + \Delta y^2) + (\Delta z^2 + \epsilon^2)$. This procedure leads to an anisotropy in the force because the $z$ direction is treated differently from $x$ and $y$. This anisotropy is accentuated because these sums are the dominant source of error within GRAPE: they are done via lookup tables in a logarithmic format with a random error of $\pm 3\%$ for each add. Figure 3 shows that the distributions of errors in the $x$ and $y$ directions are virtually identical while those in the $z$ direction are significantly different. At a given particle separation, the error will depend on the relative importance of $x$ and $y$ on one hand, and of $z$ and $\epsilon$ on the other hand. In practice, one can circumvent this anisotropy by rotating the axes that one feeds into the $x$, $y$, and $z$ inputs of GRAPE between each timestep.

### 2.3.2. Potential Energy

In addition to returning the acceleration of each particle, GRAPE returns their potential energy. Likewise, these values are not perfectly accurate. Most of the roundoff error is reduced by a factor of three relative to the force calculation because the potential uses the square root of $(\Delta r^2 + \epsilon^2)$, instead of the 3/2 power. However, since the home particle typically also appears as a neighbor particle, a self interaction contribution of $1/\epsilon$ will be summed in the potential calculation and must be removed on the host. If $\epsilon$ is very small compared to the local interparticle separation, the self interaction term can dominate the potential and roundoff error can be significant. Overflows can occur if scaling parameters are poorly chosen: $M > 64\, m\, \epsilon/d$, where $M$ is the mass of a particle and $m$ and $d$ the mass and distance scales, respectively. Cases with significant roundoff error will be in very low density regions whose contribution to the total potential energy of a simulation is not important.

TABLE 2
GRAPE FORCE ERRORS AS A FUNCTION OF PARTICLE NUMBER

| | Error (%) | |
| $N_{part}$ | Mean | r.m.s. |
| --- | --- | --- |
| 1 | 0.06 | 2.0 |
| 3 | 0.09 | 1.5 |
| 10 | 0.11 | 1.2 |
| 30 | 0.10 | 1.1 |
| 100 | 0.09 | 1.0 |
| 300 | 0.09 | 0.8 |
| 1,000 | 0.09 | 0.5 |

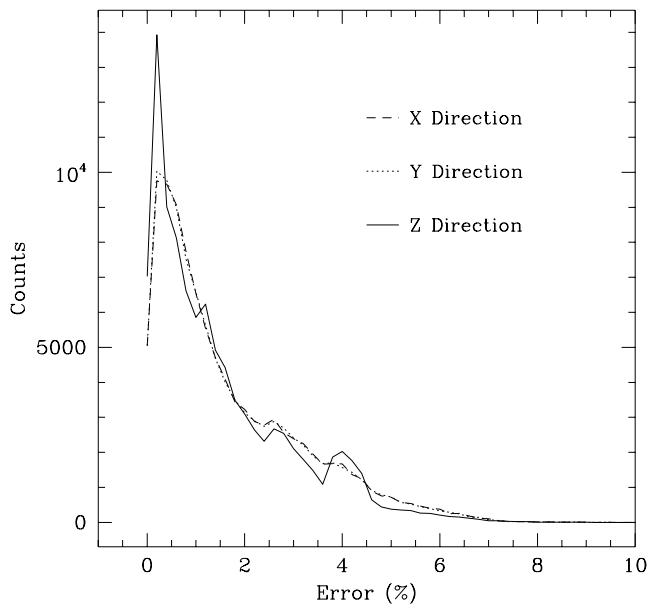

FIG. 3.—Distribution of errors (absolute value) in the directional forces returned by GRAPE for randomly separated pairs of particles.

### 2.3.3. Neighbor Lists

The last piece of information provided by GRAPE is a list of neighboring particles (useful for, e.g., Smoothed Particle Hydrodynamics). These are found by compiling the list of particles satisfying the equation $\Delta r^2 + \epsilon^2 < s^2$, where $s^2$ is the search parameter supplied by the user. Note that the radius of the sphere that is searched, $r_s$, is related to the search parameter by $s^2 = r_s^2 + \epsilon^2$. If the number of neighbors returned by GRAPE is equal to the maximum allowed (256 here), the list has overflowed and the user has to call a new calculation with a reduced search parameter for the list to be meaningful. Due to roundoff errors, GRAPE will return neighbors that are slightly outside the search sphere or, inversely, not return some that are just inside. This problem only affects of the order of one particle out of a few thousand for a random distribution. It can be solved by specifying a search parameter about 8% larger than actually desired and then sorting the list on the host (Makino 1993, private communication; see also Summers 1995).

### 2.4. Astrophysical Test Problem for GRAPE

In light of the inherent errors discussed above, it is instructive to see how the GRAPE hardware handles the least complex of astrophysical problems: the two body problem. This problem is exactly the type of problem for which the GRAPE–3 board was *not* designed; detailed orbits are better addressed by the higher accuracy GRAPE–2 boards. The GRAPE–3 hardware is geared toward many body collisionless systems where the errors on a given pair of particles can be significant, but average out over many pairs.

We placed two equal mass particles on a circular orbit and monitored the evolution of eccentricity and radius of the orbit over 14 dynamical times. The calculation was done with both direct summation and the GRAPE libraries. For direct summation, a constant timestep of $10^{-4}$ (in units of dynamical time) gave an almost perfect evolution; one of $10^{-3}$ gave an error of less than 0.5%. Using GRAPE and a leapfrog integrator with constant timesteps gave a diverging orbit for both of these values after a few dynamical times. Instead, we implemented a predictor–corrector scheme, where the timestep is updated at each step according to criteria based on accelerations and their variation, as well as on the softening parameter used for the particles. Figure 4 illustrates how the radius of the orbit oscillates within 2% and how its eccentricity stays below 0.02. Other conserved quantities (energy and momenta) were also stable, although errors were somewhat larger (from 5 to 10%). Thus, with a careful selection of timesteps, GRAPE–3A can follow orbits accurate to a few percent.

This test indicates that it should be possible to use GRAPE for N–body simulations despite its intrinsic inaccuracy. Provided sufficient care is taken in the selection of GRAPE parameters and simulation techniques, one can achieve reasonably low errors while taking advantage of its very high computational speed. In cosmological simulations, many particles will interact with a given particle and the cumulative effect will mitigate the pairwise error discussed above. For such simulations, one must be careful not to put too much emphasis on individual orbit analysis.

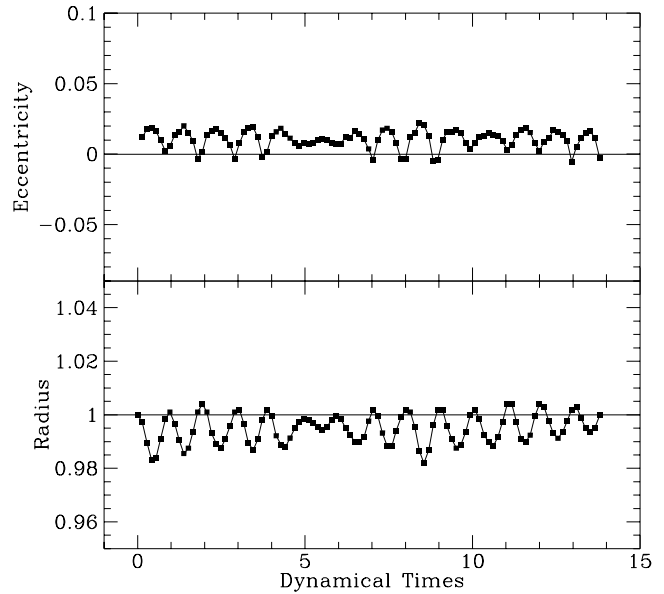

Fig. 4.—Evolution of the eccentricity and radius of a circular two body orbit. Initial values are 0 and 1, respectively. (Negative eccentricities correspond to roundoff errors.)

### 3. P3M ON GRAPE

#### 3.1. Principle of $P^3M$

The $P^3M$ method was introduced to numerical cosmology by Efstathiou & Eastwood (1981) as a refinement of the PM grid based technique (see HE). The PM calculation uses fast Fourier transforms (FFTs) to compute the potential from the density field smoothed onto a regular grid and then interpolates the forces at particle positions. The resolution of PM codes is limited by the grid spacing. The $P^3M$ method treats the PM calculation as a long range force and adds a PP calculation as a short range force to correct the forces down to a user specified gravitational softening scale. The combination of techniques has improved resolution over pure PM codes and is much faster than pure PP codes because the direct summation calculation is done only out to a limited radius (called the cutoff radius, $r_c$). Figure 5 illustrates how the combination of PM and PP forces achieves a Plummer force (the arrow indicates $r_c$).

In discussing the GRAPE implementation of $P^3M$, it helps to understand some of the details of the PP calculation. The simulation volume is divided into cubical cells with a side length greater than or equal to the cutoff radius (this "chaining mesh" is distinct from the PM grid). To guarantee the correct short range force, each particle in a cell must have a pairwise interaction calculated for every particle within its cell and in the 26 surrounding cells. Those pairs separated by more than the cutoff radius are assigned zero force. Below that cutoff radius, the PP force is tabulated as the difference between the total force and the PM force (as a function of particle separation) and stored in a lookup table. One should note that pairwise interactions will be calculated not only for all pairs with separations less than $r_c$, but also for some pairs with separations between $r_c$ and $2\sqrt{3}\,r_c$.

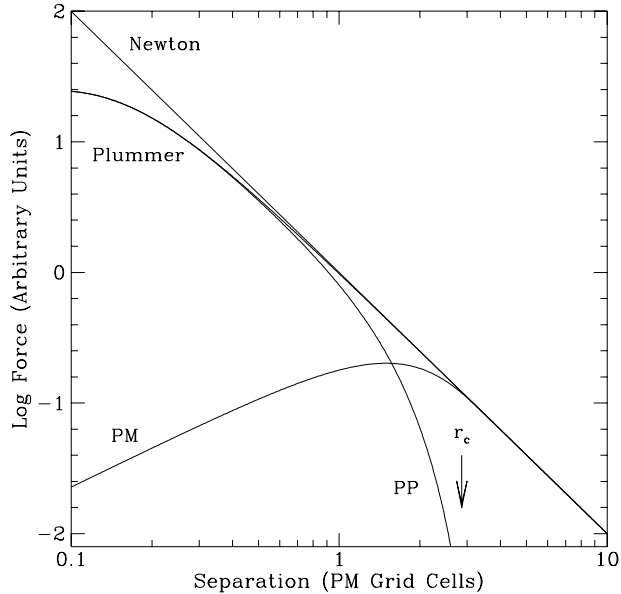

FIG. 5.—Decomposition of the P$^3$M force law. The straight line labeled "Newton" represents an inverse square law. The Plummer law is the same, except on small scales where the softening (1/8 of a cell) is significant. The PP force follows the Plummer law on small scales, but drops off sharply to 0 at the cutoff radius ($r_c$). By construction, the PP force is the difference between the Plummer and PM forces.

### 3.2. P3MG3A Algorithm

The P$^3$M algorithm becomes slow when clustering develops, as more and more pairwise interactions must be computed. Hence, the basic idea is to use GRAPE for the PP calculations. However, the GRAPE board has a hard wired Plummer force law and the desired PP force in P$^3$M is something quite different (see Figure 5). Also, it is not possible for the GRAPE board to zero out interactions that are beyond the cutoff radius as is done in standard P$^3$M. Although using GRAPE in a pure direct summation code (as was done previously) encounters neither of these problems, such calculations are limited to a few $10^5$ particles on our hardware (see Section 2.2). Therefore, some modification of the PM and PP combination is needed to handle larger simulations. We call this code P3MG3A for P$^3$M on GRAPE-3A.

The main problem is that, for the interactions which the board handles, one call to GRAPE returns the "correct" total force, not the total force minus the PM force as desired. One needs an efficient way of subtracting out the PM force for those interactions and those interactions only. Several methods were investigated, but we found that the most efficient is to call GRAPE several times. The first call returns the desired total force (Plummer law) and the subsequent calls approximate the negative of the PM force. Since GRAPE can return Plummer laws only, the idea is to find a combination of such laws (with varying softening lengths) that behaves like the PM force, in order to cancel it by subtraction. We call this the canceling force:

$$F_{\rm PM} \simeq -F_{\rm cancel} = \sum_i \alpha_i \frac{r}{(r^2 + \epsilon_i^2)^{3/2}},$$

where the masses and gravitational constant are taken as unity for convenience. The total force is:

$$F_{\rm total} = F_{\rm Plummer} + F_{\rm PM} + F_{\rm cancel}$$

where $F_{\rm Plummer}$ is the first GRAPE call, $F_{\rm cancel}$ is the sum of the subsequent GRAPE calls, and both are local calculations only. This way we take advantage of the speed of the FFTs for the long range interactions and of the speed of GRAPE for the short range interactions.

A few other points on the algorithm should be made. First, because GRAPE returns only the sum of all pairwise interactions for the home particle, symmetry of forces (i.e. Newton's second law) cannot be used as in standard P$^3$M to speed up the PP calculation by a factor of two. Second, to avoid communication overhead one can use direct summation on the host instead of GRAPE when the number of interactions to calculate is small. We found that the break even point was when the number of interactions, i.e. the number of home particles times the number of neighbor particles, is around 3,000. Finally, the potential energy calculation is done in a manner analogous to that given above; the same parameters that provide a good fit to PM force also provide a good fit to the PM potential.

### 3.3. Force Law

The analytical approximation to the PM force used in our code is given in Section 8-7-5 of HE. We have searched for fits to this expression with one and more Plummer laws over the range on which forces are computed by direct summation. For fiducial values of $a = 3.83$ PM grid cells and $r_c = 0.75\,a$ (see HE, eq. 8-72 and 8-73), the important range is from 0.1 to 10 PM grid cells. The algorithm checks that at each of the sample points (on both linear and logarithmic scales) the error in the force is lower than the maximum we allow, and then finds the values of $\alpha$ and $\epsilon$ that give the best least squares fit.

Unfortunately, although similar, the PM and Plummer forces are not exactly the same. The PM force turns over more sharply than the Plummer law (see Figure 6) and a single extra call to GRAPE will not suffice. One could reshape the PM force to match the Plummer law, but the slow turnover of a Plummer law would force the cutoff radius to be excessively large. If we want the PM force to approximate a Plummer law with a softening of 1 PM grid cell, then to achieve less than 2% discontinuity at the cutoff radius, one must increase $r_c$ from 2.9 to 8.6 PM grid cells. The number of pairwise interaction calculations would increase dramatically and overwhelm the gain from fewer calls to GRAPE.

We found that two Plummer laws can match the PM force reasonably well and that using more than two does not improve the fit enough to justify the expense of additional calls to GRAPE. After searching a constrained four dimensional parameter space (weight $\alpha_i$ and softening $\epsilon_i$ of each Plummer law), we found several fits that give the correct force to within 5% on all relevant scales. This result compares well with the value of 6% for the Adaptive P³M code (Couchman 1991). Because there are then three calls to GRAPE for each force calculation, the maximum speedup over direct summation one could expect (100) is divided by three: we cannot hope for more than a factor of 30.

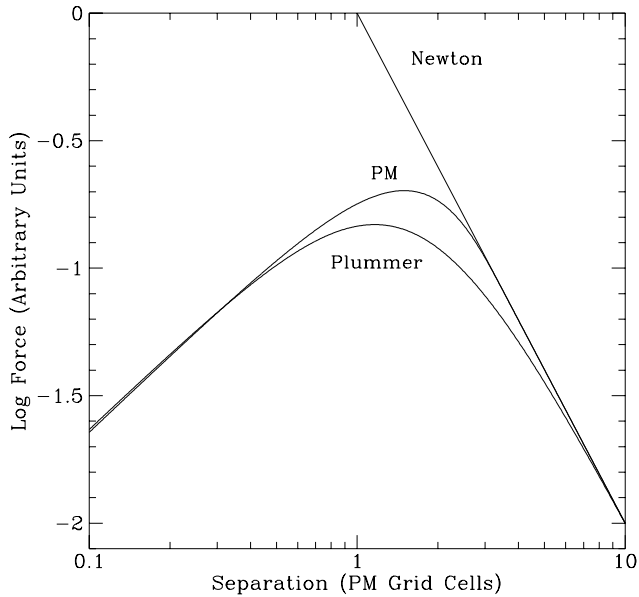

Fig. 6.—Comparison of force shapes: PM force from the previous figure and Plummer law with softening of 1.6 PM grid cells.

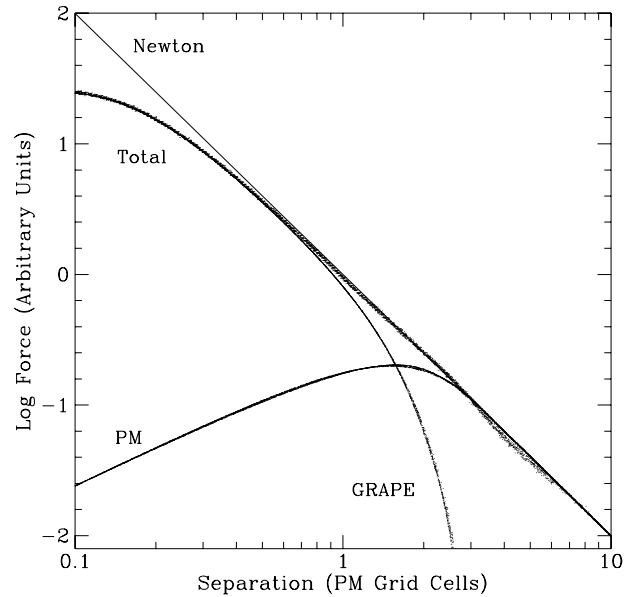

Fig. 7.—Decomposition of the P3MG3A force law — same format as Figure 5. Points represent forces returned by the code for 10,000 pairs of particles with random logarithmic separations. The curve labeled "GRAPE" is the sum of $F_{\text{Plummer}}$ and $F_{\text{cancel}}$ (see text).

The force scatter inherent to GRAPE is amplified by the weighting of Plummer laws used to cancel the PM force. The scatter in the PM force is reduced to less than 2% by the use of the QPM method (see HE, Section 9-1-3), so it is desirable to keep the scatter in the PP forces down to a reasonable level ($\leq 5\%$). This in turn forces the use of small weights, which makes it more difficult to obtain a very good fit to the PM force. We believe we have found a satisfactory compromise with the following fit:

$$F_{\text{cancel}} = -2.4 \frac{r}{(r^2 + 1.9^2)^{3/2}} + 1.6 \frac{r}{(r^2 + 3.5^2)^{3/2}},$$

where distances are in units of PM grid cells. Naively, one would expect the sum of the scaling factors ($\alpha_i$) to equal $-1$ in order to exactly cancel the PM force as $r \to \infty$. However, with such large softening values, this regime is not encountered until well outside the largest scale of the PP calculation. The resulting force law is shown in Figure 7 including the errors due to GRAPE. It is quite accurate, except around 5 PM grid cells, where it is slightly underestimated. We expect that errors in this intermediate regime will not be very important in cosmological simulations because the major influences come from either local non linear structures or tidal forces at large distances, for both of which the forces are correct. In the regime from $r_c$ to $2\sqrt{3}\,r_c$ (i.e. 3.9 to 13.4 PM grid cells here) the force is calculated by either pure PM force *or* by the PM plus GRAPE forces, depending on the pair placement relative to the chaining mesh — which makes reshaping the PM force an impractical option.

### 3.4. Technical Issues

We use the P3MSPH code described in Summers (1993). This code was based on that of Evrard (1988), which derived its P³M part from Efstathiou et al. (1985). Calls to GRAPE were added as described above and the SPH part turned off at this point. Thanks to GRAPE's speed we are more memory than CPU limited. Under our current 160 megabyte configuration, we can handle a $128^3$ particle simulation on a $256^3$ cell grid at best.

We have performed some timing estimates for such simulations on the host with and without GRAPE, and on a single processor of a Cray C-90. These times were averaged over three steps when the particle distribution is highly clustered. Results are shown in Table 3. CPU times are in minutes for various hardware with speedup factors relative to P³M on a SPARCstation 10/51 in brackets where relevant. The FFT is faster on the Cray C-90 because it is vectorized, but not for P3MG3A, since it is performed on the host regardless of whether the board is used. The PP force calculation on GRAPE is two thirds of the maximum speedup. An important point is that the P3MG3A calculation is now well balanced between the PM part and the PP part, while the PP part is heavily dominant in standard P³M. The factor of ten speedup achieved will naturally be decreased for smaller runs or less clustered states. The addition of GRAPE makes large simulations feasible on a workstation and achieves almost one third the speed of one of the world's fastest single CPUs.

TABLE 3
Timings for a Single Timestep on Various Hardware

| Code<br>Hardware | CPU Time (min.) [Speedup Factor] | | |
|---|---|---|---|
| | $P^3M$<br>SPARC 10/51 | P3MG3A<br>GRAPE-3A | $P^3M$<br>Cray C-90 |
| PM Force | 8.83 | 8.85 | 1.50 [5.9] |
| PP Force | 159.88 | 7.66 [20.9] | 3.28 [48.8] |
| Miscellaneous | 0.22 | 0.22 | 0.06 |
| Total | 168.93 | 16.73 [10.1] | 4.84 [34.9] |

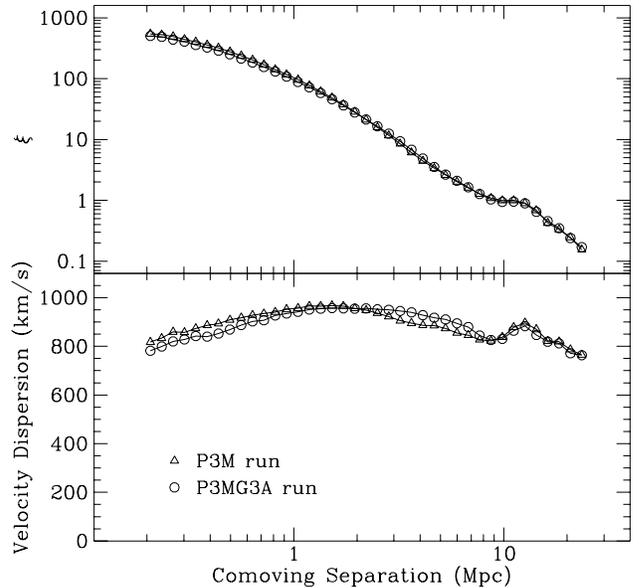

Fig. 8.—Two point correlation function $\xi$ and pairwise velocity dispersions for a test simulation. (See text for details.)

To extrapolate the CPU time required for a complete run, one must take into account that the $P^3M$ timings are highly dependent on the evolution of clustering (much less so for P3MG3A). Based on previous experience, reasonable estimates for a similar simulation with 1,000 timesteps are 1.2 CPU day on the C-90, 10.7 on GRAPE, and 39 on the host without GRAPE. Actually, in terms of wall clock rather than CPU time, it is likely to take less time to run on GRAPE than on a supercomputer. Considering the time associated with file transfer, waiting in batch queues, and time shared systems, an eleven day turnaround at a supercomputer center is quite good. Furthermore, there is no need to apply for computing time and no external restrictions on running times or maintenance. The total cost of our setup is around USD 45,000.

## 4. TESTS AND SIMULATION

Having shown that the speed of GRAPE makes the code well worth pursuing, we now look for liabilities associated with the force inaccuracies. As a first test of the P3MG3A code we have run a simulation of a COBE normalized Cold Dark Matter universe using $32^3 = 32{,}768$ particles and $64^3 = 262{,}144$ PM grid cells. The volume is a cube 100 Mpc across and the gravitational softening is set at 195 kpc. The simulation was evolved from redshifts $z = 20$ to $z = 0$ in 1,000 timesteps. Comparative runs were done on the workstation with and without GRAPE, on a Cray C-90, and with a staggered mesh pure PM code (Cen 1992) using a $256^3$ PM grid. Since standard $P^3M$ is known to agree with other codes used in cosmology (Weinberg at al. unpublished), this allows a fair assessment of the reliability of GRAPE. Results of comparisons with other codes will be reported in a separate paper (Cen et al. 1995).

In a simple visual analysis, differences are quite hard to find. We have produced a 50 frame 3D motion picture of the simulations showing the particles color coded with their velocities in a sliced rotating box (Brieu 1994). The visualizations of the $P^3M$ and P3MG3A runs look remarkably identical. Energy conservation, as measured by the Layzer–Irvine equation, is estimated at the 0.5% level for the P3MG3A run compared to about 0.2% for the other runs. The GRAPE value is only an estimate because the potential energy returned from the board contains roundoff errors up to a few percent and the integral deviates by random walk even if energy is being exactly conserved. Reasonable conservation is achieved because the potential energy is dominated by close pairs in high density regions where the GRAPE errors are relatively small.

Basic cosmological statistics indicate some small variations. Figure 8 shows the final two point correlation function $\xi$ and pairwise velocity dispersions for both simulations. Correlations are close to indistinguishable on all scales, though the GRAPE run tends to be marginally lower at small separations. Velocity dispersions show a larger disagreement, but still less than 10%. The three dimensional nature of velocity dispersions make them subject to more variation than the one dimensional correlations. Differences occur below the non linear scale of the simulation and seem to indicate that the P3MG3A run has not relaxed down to small scales as much as the $P^3M$ run.

Another way to get a handle on particle clustering is via the group multiplicity function. Here we apply the friends of friends grouping algorithm (Davis et al. 1985) with various linking lengths and tabulate the number of groups as a function of the number of particles in a group. The linking length is traditionally specified as $\eta L/N$, where $\eta$ is the linking length parameter and $L/N$ is taken to be the mean separation between particles for a box of side length $L$ containing $N^3$ particles. Results are shown in Figure 9. Differences are minor except in the densest regions: large groups found with a small linking length are somewhat more abundant in the $P^3M$ run. However, since at $\eta = 0.05$ the linking length is smaller than the gravitational softening scale, little should be made of this discrepancy but to say that it agrees in character with that of the velocity dispersions above.

More evidence can be found in the comparison of density fields. Figure 10 compares the density smoothed onto a $64^3$ grid via the TSC method (HE, Chapter 5). The ratios of densities in a cell by cell comparison were divided into logarithmic bins and the mean and standard deviations in each bin are plotted. The Cray versus SPARC $P^3M$ runs show the negligible scatter induced simply by roundoff error (8 byte words versus 4 byte words). The PM versus $P^3M$ comparison typifies the deviations one may expect between different codes, though one should note that the PM run had a nominal resolution a factor of two larger than the other runs. The GRAPE run shows the now expected underestimation of densities at the high end plus a compensating overestimation at intermediate values. The trends are the same as for the PM code, only stronger at high densities and weaker at low densities. Considering its chaotic nature, this comparison can only be used to make sure that there is no major discrepancy (Weinberg et al. unpublished). The fact that deviations are kept mainly within 10% indicates quite reasonable results.

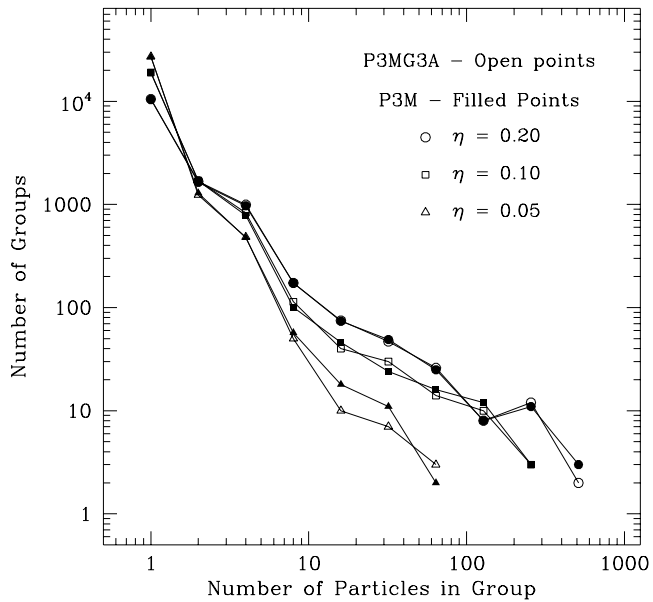

Fig. 9.—Group multiplicity functions for the $P^3M$ and P3MG3A simulations for three values of the linking length. (See text for details.)

A consistent picture emerges from the previous observations: the densest regions in the P3MG3A run are slightly less evolved than those in the $P^3M$ run. The number of very high peaks as well as their densities and velocities are a bit lower in the GRAPE simulation. It appears that the small underestimation of the force during collapse prolongs the formation time and produces slightly less structure. These differences appear at the 10% level for a completely evolved simulation. In cosmology, lax observational constraints and cosmic variance will render such differences immaterial. If fine lines need be drawn, one should keep in mind that the code will tend to mildly underestimate peaks in structure development or use a different simulation technique. The P3MG3A code will have to be used with cautious knowledge of its small inaccuracies, but there can be no doubt that it is an efficient tool for cosmology.

## 5. DISCUSSION

We have demonstrated that the GRAPE special purpose hardware represents an attractive alternative to supercomputers for large cosmological simulations. Our code handles cosmological simulations with a peak efficiency of ten times a SPARC 10 and one third of a Cray C-90. A comparison of simulation results shows that the inherent pairwise force errors do not produce important deviations in the measured cosmological statistics. The hardware, the code, and a bit of prudence will enable local workstations to approach the realm of supercomputing.

Although the performance of the P3MG3A code is already impressive, two improvements can be made rather easily. First, the GRAPE board we use was the first commercial release. The latest board has twice as many chips, can handle four times as many particles per call, and returns a neighbor list four times as long. The increased throughput should lead to a quantum jump in speed. Second, the use of parallel computing can provide a quick doubling of speed. Since the P3MG3A code is well balanced between the PM and PP calculations, one can spawn off the PM part to another CPU (via a message passing interface like PVM) with relatively little penalty. Both improvements are straightforward to implement and offer a substantial increase in speed.

Other improvements are planned. The current code handles only collisionless matter and allows us to study cosmological problems where dark matter dominates, such as halos of clusters of galaxies (Summers et al. 1995). We are pursuing the addition of Smoothed Particle Hydrodynamics to the code to treat the baryonic component of the universe as well (Summers 1995). To improve the accuracy of the gravitational force law, one may search for a better fit for the canceling force, but it is doubtful that much improvement can be gained. The preferable solution would be for a GRAPE board with a user programmable force law and slightly higher accuracy. Such boards already exist, but only as prototypes.

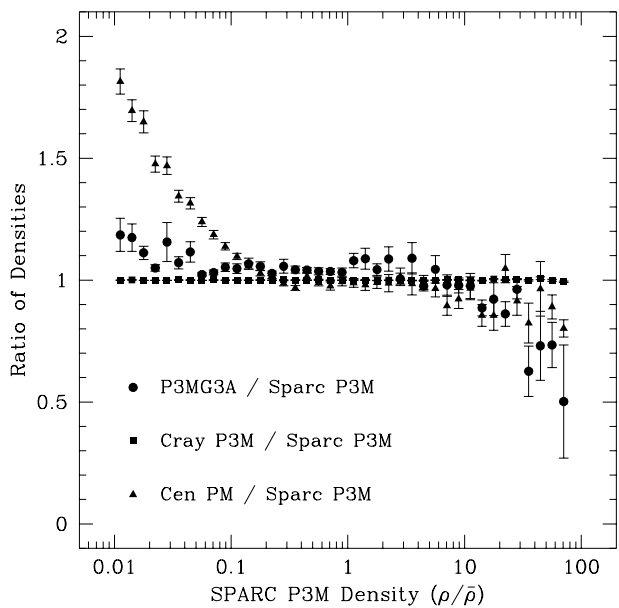

Fig. 10.—Comparison of smoothed density fields. The density fields of the test simulations were smoothed onto a $64^3$ grid with the Triangular Shaped Cloud method. Points plot the mean and error bars the standard deviation of the density ratios, relative to the $P^3M$ run on the SPARCstation, in logarithmic bins.

It is unlikely that special purpose hardware like GRAPE boards will supplant the use of supercomputers in large cosmological simulations. Certainly, the largest simulations will require memory and storage resources beyond the means of any workstation. However, for those runs that can be accommodated on local hardware, the GRAPE board provides a means to obtain the speed of a supercomputer and maintain the ease and access of a local machine. It seems a very good deal all around if one can relieve a portion of the burden on the oversubscribed national centers, get the same turnaround time on simulations, and do it all for a fraction of the cost of a supercomputer.

We are grateful to Junichiro Makino for his invaluable assistance and technical support, as well as to Masayuki Umemura. Discussions with RenYue Cen, Nickolay Gnedin, Piet Hut, UeLi Pen, David Weinberg, and Simon White were very useful. The first author wishes to thank Gus Evrard and Gary Mamon — who was his remote advisor. And many thanks are due to Edmund Bertschinger for his generous advice.

We acknowledge the Grand Challenge Cosmology Consortium (GC$^3$) and grants of computer time at the San Diego and Pittsburgh Supercomputer Centers. Some visualization was done using software from the Laboratory for Computational Astrophysics at the National Center for Supercomputing Applications. We also used Lars Hernquist's FORTRAN treecode, and thank him for his hospitality at Santa Cruz, where part of this work was done.

This work was supported by NSF grants ASC 93–18185 and AST 91–08103 and MESR grant ADR 92–343.